\documentclass{iopart}
\usepackage{graphicx}

\usepackage{color}
\usepackage{bm}
\usepackage{epstopdf}

\graphicspath{{figs2/}}

\begin{document}

\title{Functional renormalization group approach to the singlet-triplet
transition in quantum dots}
\author{E.B. Magnusson$^1$, N. Hasselmann$^{2}$ \footnote{Formerly at the International Institute of Physics, Natal, Brazil.}
, I.A. Shelykh$^{1,3}$}

\address{$^1$Science Institute, University of Iceland, Dunhagi 3, IS-107, Reykjavik, Iceland}
\address{$^2$Max-Planck-Institute for Solid State Research, Heisenbergstr.~1, D-70569 Stuttgart, Germany}
\address{$^3$Division of Physics and Applied Physics, Nanyang Technological University, 637371, Singapore }

\begin{abstract}
We present a functional renormalization group approach to the
zero bias transport properties of a
quantum dot with two different orbitals and in
presence of Hund's coupling. Tuning the energy separation
of the orbital states, the quantum dot can be driven through
a singlet-triplet transition. Our approach, based on
the approach by Karrasch {\em et al.}~\cite{Meden06} which we apply to
spin-dependent interactions, recovers the key characteristics
of the quantum dot transport properties
with very little numerical effort. We present results on the
conductance in the vicinity of the transition and compare our
results both with previous numerical renormalization group
results and with predictions of the perturbative renormalization
group.

\end{abstract}
\pacs{73.63.Kv, 73.23.Hk, 73.21.La, 73.23.-b, 71.70.Gm}
\maketitle
\section{Introduction}

Quantum dot systems are ideal laboratories to study quantum-impurity models
and allow to investigate many-body effects which result from the coupling
of the quantum dot to a Fermi bath in the leads. If
the ground state of the isolated quantum dot is degenerate, the coupling to
the Fermi sea of the leads can induce a coupling
of these different isolated ground states via virtual transitions
and can yield a new strongly entangled many body state. A famous
example is the $S=1/2$ Kondo effect in which the leads completely screen
the spin on the quantum dot and perfect transmission across the dot
is achieved.\cite{Glazman88} Kondo physics have been suggested to play an important role in
a variety of the mesoscopic transport phenomena, in particular in formation of
the famous "0.7 anomaly", where the quantized ballistic conductance through a
quantum point contact acquires an additional conductance step at $G=0.7G_0$, $G_0$
being the quantum of conductance. \cite{Meir02}
The $S=1/2$ Kondo effect can arise, if an odd number
of electrons are trapped on the dot.
If an even number is trapped,
a degenerate ground state can arise if a Hund's coupling leads to
a finite spin of the ground state.
If we consider a dot with two orbitals, separated by an energy
difference $\delta$, and occupied by two electrons in the ground
state, then a transition from a non-degenerate spin-singlet
to a threefold degenerate spin-triplet ground state can be
achieved by adjusting the energy difference of the two orbitals
and/or the Hund's coupling strength. This singlet-triplet
transition has been studied with a variety of techniques and
can be detected directly by a peak of the finite temperature
conductance at the transition
point.\cite{Sasaki01,Kouwenhoven01} For a review on the transport properties of two-electron quantum dots, see e.g.~\cite{Florens11}. 

A robust and reliable technique for studying few level systems, such as quantum dots or nanotubes,
is the numerical renormalization group (NRG) technique,\cite{Bulla08,Anders08}
which has also been applied in investigations of the singlet-triplet
transition.\cite{Izumida01,Hofstetter02,Hofstetter04} It is however
limited to relatively simple geometries since NRG calculations for
systems with more than two dots are numerically prohibitive.

An alternative approach is the functional renormalization group
(FRG).
At the moment, available truncation schemes of the FRG
equations allow for an accurate calculation of the
dynamical properties of
the single impurity Anderson model only at weak coupling,
see Refs.~\cite{Karrasch08,Bartosch09,Isidori10}.
In the limit of
a vanishing bias voltage and at zero temperature, a calculation of
the transport through the quantum dot requires however only
knowledge of the local Green's function at the Fermi energy of
the leads. A rather simple truncation scheme of the
FRG was recently proposed in Ref.~\cite{Meden06}, which,
despite its simplicity, is accurate not just at weak coupling
but also at moderate and even at rather strong coupling.
The advantage of this approach, which we refer to as the
static approximation since the frequency dependence
of all vertex functions is neglected, is
that it is considerably faster than NRG, and
also simpler than other alternative approaches such as
Fluctuation Exchange Approximation (FLEX) \cite{Horvath10} or field theoretical RG approaches \cite{Hermann}
and can be easily applied
also to multidot geometries.\cite{Meden06}
The approach in Ref.~\cite{Meden06} was devoted
to the study of density-density interactions and a
Hund's exchange was not discussed.
Here, we use the approach to investigate QD's with Hund's exchange.
We present results
for a two level QD with Hund's coupling and compare our results
with previous works.

\section{The model}
Considering a QD with two orbitals, there are multiple interactions to
account for. Firstly,
there are two intra- and one inter-orbital density-density
interaction terms.  Secondly, a Hund's  spin interaction can exist
between the two orbitals (in the static approximation, there is no independent
intra-orbital spin interaction since for $S=1/2$ it can always be written
as a density-density interaction).
We thus arrive at the Hamiltonian of the isolated dot of the form
\begin{eqnarray}
  H_d &= \sum_{\sigma} \left[
    E_{A\sigma} a_{A\sigma}^\dag a_{A\sigma} +E_{B\sigma} a_{B\sigma}^\dag
    a_{B\sigma} \right]
  \nonumber \\
  & \quad  + U_{AB} \big(n_{A}-1) \big(n_{B}-1)
  + U_A \big(n_{A\uparrow}-\frac{1}{2}\big)\big(n_{A\downarrow}-
  \frac{1}{2}\big) \nonumber
  \\ & \quad
  +U_B \big(n_{B\uparrow}-\frac{1}{2}\big) \big(n_{B\downarrow}-
  \frac{1}{2}\big)
  \nonumber \\
  & \quad
  + \frac{J}{4} \sum_{\sigma_1,\sigma_2,\sigma_3,\sigma_4}
  {\bm \sigma}_{\sigma_1\sigma_2} \cdot{\bm\sigma}_{\sigma_3\sigma_4} \,
  a^\dag_{A\sigma_1} a_{A\sigma_2} a^\dag_{B\sigma_3}  a_{B\sigma_4}   \, ,
  \label{eq:Hd}
\end{eqnarray}
where $U_A$ and $U_B$ are intra-orbital interaction energies, $U_{AB}$ the inter-orbital
interaction energy and $J$ is the Hund's coupling energy. Here, $\bm \sigma=(\sigma^x, \sigma^y, \sigma^z)$
are the Pauli matrices with elements ${\bm \sigma}_{\sigma_1\sigma_2}$.
In the presence of a small magnetic field $H$ the energy levels are Zeeman-split,
\numparts
  \begin{eqnarray}
    E_{A\sigma}& = E_A + \sigma h \, ,
    \\
    E_{B\sigma}& = E_B + \sigma h \, ,
  \end{eqnarray}
  with $\sigma=\pm 1$, $h=g \mu_B H/2$ ($\mu_B$ is the Bohr-magneton and $g$ the electron $g$ factor) and
  \begin{eqnarray}
    E_A & = V_g-\delta/2 \, , \\
    E_B & = V_g+\delta/2 \, ,
  \end{eqnarray}
  \label{eq:levpos}
\endnumparts
where $V_g$ is the gate voltage
and $\delta$ is the difference of the two orbital energy levels.
Note that $V_g=0$ corresponds to half filling of the dot, as we put the Fermi energy of the leads to be zero.
The leads are described by
\begin{eqnarray}
  H_R+H_L&=\sum_{\tau\sigma k}\varepsilon_k \left(  c_{R\tau\sigma k}^\dag
c_{R\tau\sigma k}+ c_{L\tau\sigma k}^\dag c_{L\tau \sigma k}\right) \, ,
\end{eqnarray}
where $k$ labels the momenta in the leads and $\tau=A,B$ denotes the different
orbital quantum numbers.
The coupling of the leads to the dot is described by the tunneling
term
\begin{eqnarray}
  H_T &=\sum_{k\tau\sigma} \left(t_{R\tau} c_{R\tau\sigma k}^\dag
    a_{\tau\sigma}+t_{L\tau}c_{L\tau\sigma k}^\dag a_{\tau\sigma} +
    \rm{ h.c.}\right) \, .
  \label{eq:tunnel}
\end{eqnarray}
Note that the hopping
conserves the orbital quantum number $\tau=A,B$, which makes the
model appropriate for vertical
coupled dots.
In contrast, in lateral quantum dots, the orbital quantum number
is not conserved. Lateral dots are described by a different tunneling
term of the form
$H_T =\sum_{k\sigma\tau} \left(t_{R} c_{R\sigma k}^\dag a_{\tau\sigma}+t_{L} c_{L\sigma k}^\dag a_{\tau\sigma} + \rm{h.c.}\right)$ and both orbitals share the  bath electrons $H_R+H_L=\sum_{\sigma k}\varepsilon_k \left(  c_{R\sigma k}^\dag
c_{R\sigma k}+ c_{L\sigma k}^\dag c_{L\sigma k}\right)$.
The absence of the orbital quantum number in the leads allows for virtual processes where the
$\tau$ quantum number of the dot electrons are changed.
Later quantum dots
where studied e.g.~in
Refs.~\cite{Izumida01,Hofstetter02,Pustilnik03,Hofstetter04}.
Vertical quantum dots have been investigated by Pustilnik and
Glazman \cite{Pustilnik01} using scaling techniques and
Izumida and Sakai \cite{Izumida01} using NRG.

\subsection*{Conductance calculation}
Since the tunneling Hamiltonian Eq.~(\ref{eq:tunnel}) conserves the orbital
quantum number and the spin projection along the magnetic field,
the total (zero bias) conductance through the dot is the sum of the partial conductances
through each orbital level and spin projection.
These can be obtained from the single-particle
Green's function using the Landauer-B\"uttiker formula,\cite{Meir92}
\begin{eqnarray}
  G & = \sum_{\tau\sigma} G_{\tau\sigma} \, ,
  \label{eq:G1}
\end{eqnarray}
with
\begin{eqnarray}
  G_{\tau \sigma} =\frac{e^2}{\hbar} \rho_{\tau \sigma} \Gamma_{L \tau} \Gamma_{R \tau} / (\Gamma_{L \tau} + \Gamma_{R \tau})\, .
  \label{eq:conductance}
\end{eqnarray}
Here, $\rho_{\tau\sigma}(0)$ is the density of states of the quantum
dot at the
Fermi energy and
$\Gamma_{\tau}^{L/R}=2\pi |t_\tau^{L/R}|^2 \rho_{0}$ is a broadening of the level with $\rho_0$
being the density of states in the leads.
The dot density of states can be obtained from the
Fourier-transform of the retarded Green's function
of the dot in presence of the leads,
${\cal G}_{\tau\sigma}(t-t')= -i\theta(t-t') \langle [a_{\tau\sigma} (t) , a_{\tau\sigma}^\dagger (t')]\rangle $,
\begin{equation}
  \rho_{\tau\sigma}(\varepsilon) = -\frac{1}{\pi} \rm{Im} \mathcal {\cal G}_{\tau\sigma}
(\varepsilon+i0^+) \, .
\label{eq:G3}
\end{equation}

\section{Singlet-Triplet transition}

Here we concentrate on the situation with $U_A=U_B=U_{AB}$ so that,
for total occupancy $N=\sum_{\tau \sigma} n_{\tau\sigma}=2$,
there is no capacitative energy associated with transferring an electron
from orbital $A$ to orbital $B$. There is however a total energy
cost which arises from both the Hund's coupling and from the
energy difference of the orbitals. If both electrons occupy the lower
$A$ orbital, they necessarily form a (local) singlet. On the other hand,
if one electron occupies the lower $A$ orbital and another one the
upper $B$ orbital, then for
$J<0$ the lowest energy can be achieved for a ferromagnetic spin alignment
of the two electrons. The difference between the local singlet and
the triplet state is
\begin{equation}
  K_0= \delta-J/4 \, ,
  \label{eq:STcondition}
\end{equation}
and we thus expect a singlet-triplet transition at $K_0=0$, where
the configuration changes from a non-degenerate singlet to a
threefold degenerate triplet state. In reality, because of
renormalization effects, the transition is shifted from the point
$K_0=0$. One can however
replace $K_0$ by the renormalized quantity $K$ such that transition takes place at $K=0$.

\subsection*{Symmetric coupling}
Here we focus on the symmetric situation  where
$t_{L\tau}=t_L$, $t_{R\tau}=t_R$ and thus
$t^2=t_{L\tau}^2+t_{R\tau}^2$ is independent of $\tau$.
On the triplet side of the transition the spins are locked together
in a spin-1 state and the problem becomes a two channel spin-1
Kondo problem with a completely screened groundstate
and maximal conductance
\begin{equation}
  G=\frac{4 e^2}{h} \left( \frac{2 t_L t_R}{t_L^2+t_R^2} \right)
\end{equation}
at $T=0$, while on the singlet side, conduction is effectively blocked. The analysis of Ref.~\cite{Pustilnik01}, based
on Anderson's poor-man-scaling technique,
predicts
a monotonous decrease of $G$ with $K$ on the singlet side $K>0$,

\begin{equation}
  G/G_0=B \ln^{-2} \left[K/T_0\right] \, , \, \, \,
  B=\left(\frac{3\pi}{8}\, \frac{\lambda-1}{\lambda+1} \right)
  \label{eq:asymptote}
\end{equation}
with $G_0=2e^2/h$,$\lambda=2+\sqrt{5}$ and the characteristic energy scale
\begin{equation}
  T_0\simeq U \exp[-\tau_0 \pi U/ 8 \Gamma]
\end{equation}
where $\tau_0\approx 0.36$. The asymptotic behavior (\ref{eq:asymptote})
is valid for $K\gg T_0$.
At finite temperature, the conductance develops a peak near
$K\approx 0$. Both the asymptotic behavior and the peak near $K=0$
are recovered also in the FRG approach discussed below.

\section{Functional Renormalization Group}

According to Eqs.~(\ref{eq:conductance},\ref{eq:G3})
the calculation of the conductance relies on the retarded Green's function, which we
evaluate using the Matsubara imaginary time formalism.
The full Green's function is defined as
\begin{equation}
\big[{\cal G}_{\tau\sigma}(i\varepsilon)\big]^{-1}=\big[{\cal G}_{0,\tau\sigma}(i\varepsilon)\big]^{-1}
-\Sigma_{\tau\sigma}(i\varepsilon) .
\end{equation}
where $\Sigma_{\tau\sigma}$ is the particle self-energy and ${\cal G}_{0,\tau\sigma}$ is
the noninteracting Green's function of the dot  which, after integrating out the leads, has the form
\begin{equation}
  {\cal G}_{0,\tau\sigma}(i\varepsilon)=\frac{1}{i \varepsilon
    +i \Gamma_{\tau}\mbox{sgn} (\varepsilon)- E_{\tau\sigma}} \, ,
\label{eq:G0}
\end{equation}
where $\Gamma_\tau=2\pi t_\tau^2 \rho_{0}$,
$t_\tau^2=|t_{L\tau}|^2+|t_{R\tau}|^2$ and
$\rho_0$
is density of states  in the
leads which we take to be independent of $\tau$ and $\sigma$.
All interaction effects are contained in the
self-energy and below we calculate it using the non-perturbative
FRG technique.

The functional RG approach
is based on an exact flow equation of the generating functional
of irreducible vertices which can be used to derive flow equations
for the irreducible vertices themselves.\cite{Wetterich93,Morris94,Metzner11}
The irreducible two point vertex is the self energy,
the irreducible four point vertex corresponds to the fully renormalized
two-particle interaction, and so on.
This yields a hierarchy
of coupled flow equations for all vertices which we truncate at the
four point vertex, assuming that all vertices of higher order do
not play an important role.
In the scheme we use below, both the two-point and the four-point
vertex are further assumed to be independent of energy, a rather
drastic approximation which however was shown to yield very
accurate results for the zero temperature linear conductance of
coupled quantum dots.\cite{Meden06}
In the geometry of the quantum dots studied
here, the two-point vertex is further diagonal in both orbital
and spin quantum numbers if we choose the spin quantization axis to be parallel to the direction of the magnetic field.

The flow equations emerge from introducing an infrared cutoff $\Lambda$ in the non-interacting Green's function such that ${\cal G}_0^\Lambda \rightarrow 0$ for $\Lambda \rightarrow \infty$ and ${\cal G}_0^\Lambda = {\cal G}_0$ for $\Lambda=0$.
We introduce the cutoff
via a multiplicative step-like regulator function
\begin{equation}
  {\cal G}_{0,\tau\sigma}^\Lambda(\varepsilon)
  = \theta(|\varepsilon|-\Lambda){\cal G}_{0,\tau\sigma} \, ,
  \label{eq:G0reg}
\end{equation}
where $\theta$ is the Heaviside step function. Taking the derivative of the equation for the generating functional with regard to $\Lambda$ then yields differential equations for the irreducible vertices, and integrating from $\Lambda \rightarrow \infty$ down to $\Lambda=0$ yields the fully renormalized vertices.

The flow equation
of the self-energy $\Sigma^\Lambda_1=\Sigma^\Lambda_{\tau_1\sigma_1}$
(the superscript $\Lambda$ indicates that it is the self-energy at the cutoff
value $\Lambda$, this applies to other quantities as well)
takes the form

\begin{equation}
\partial_\Lambda \Sigma^\Lambda_{1} = \sum_{2}
\int \frac{d \varepsilon}{2\pi}
e^{i \varepsilon 0^+}
\dot{\cal G}^\Lambda_{2}(\varepsilon) \gamma_\Lambda^{(4)}(1,2;2,1)
\label{eq:flowsigma}
\end{equation}
where $\gamma_\Lambda^{(4)}$ is the irreducible four-point vertex
and the lower indices $1,2$ refer to both orbital and spin quantum
number.
The single-scale propagator is defined as
\begin{equation}
  \dot{\cal G}_{1}^\Lambda(\varepsilon)=-
  \big[{\cal G}^\Lambda_{1}(\varepsilon)\big]^2  \partial_\Lambda
  \big[{\cal G}_{0,1}^\Lambda(\varepsilon) \big]^{-1}
  \, ,
\label{eq:singlescale}
\end{equation}
where
${\cal G}_{0,1}^\Lambda(\varepsilon)={\cal G}_{0,\tau_1\sigma_1}^\Lambda(\varepsilon)$, with ${\cal G}_{0,\tau_1\sigma_1}^\Lambda(\varepsilon)$ defined
in Eq.~(\ref{eq:G0reg}).

Our sharp cutoff regulator yields a simple structure for both the single-scale propagator
and the full Green's function, with
\numparts
  \begin{eqnarray}
    {\cal G}_{\tau\sigma}^\Lambda(\varepsilon)&= \frac{\Theta(|\varepsilon|-\Lambda)}
    {i \varepsilon + i \Gamma_\tau \mbox{sgn}(\varepsilon)-
      E^{\Lambda}_{\tau\sigma}}
  %{G}_{\tau\sigma}(\varepsilon)
    \, , \\
    \dot{\cal G}_{\tau\sigma}^\Lambda(\varepsilon)&= - \,
    \frac{\delta(|\varepsilon|-\Lambda)}
    {i \varepsilon + i \Gamma_\tau \mbox{sgn}(\varepsilon)-
      E^{\Lambda}_{\tau\sigma}} \, ,
  \end{eqnarray}
\endnumparts
where the initial level energy and the self-energy have been merged
\begin{equation}
  E^\Lambda_{\tau\sigma} = E_{\tau\sigma}+\Sigma_{\tau\sigma}^\Lambda \, .
\end{equation}
With the approximation of an energy-independent four-point vertex, its flow
is determined by the equation (we neglect here the contribution
of the six-point vertex),
\begin{eqnarray}
  \partial_\Lambda \gamma_\Lambda^{(4)}(1',2';2,1) &=
  -\int \frac{d \varepsilon}{2\pi}\sum_{3,4}  \Big\{
  {\cal G}_{3}^\Lambda(\varepsilon) \dot{\cal G}_{4}^\Lambda(-\varepsilon) \nonumber \\
& \qquad
\times \gamma_\Lambda^{(4)} (1',2';3,4)
\gamma_\Lambda^{(4)}(4,3;2,1) \nonumber \\
& \quad
+ 2 {\cal G}_{3}^\Lambda(\varepsilon) \dot{\cal G}_{4}^\Lambda(\varepsilon) \nonumber \\
& \qquad \times
\big[\gamma_\Lambda^{(4)} (1',3;4,2)\gamma_\Lambda^{(4)}(2',4;3,1)
\nonumber \\
& \qquad
 -\gamma_\Lambda^{(4)} (1',3;4,1) \gamma_\Lambda^{(4)} (2',4;3,2)
 \big]
\Big\} \, .
\label{eq:flow4pnt}
\end{eqnarray}
This flow equation involves products $\Theta(x) \delta(x)$ which
must be evaluated according to the prescription
$f[\Theta(x)] \delta(x)\rightarrow \delta(x) \int_0^1 dy f[y]$ with some
function $f[y]$, see Ref.~\cite{Morris94}.

With the approximation of energy-independent vertices, the conductance
contribution from each $\tau \sigma$, Eq.~(\ref{eq:conductance}),
takes a particularly simple form,
\begin{equation}
  G_{\tau\sigma}=\frac{e^2}{h}
  \frac{\Gamma_\tau^2}{\Gamma_\tau^2+\bar{E}_{\tau\sigma}^2} \, ,
\end{equation}
where we further assumed that $t_{L\tau}=t_{R\tau}$ and where
$\bar{E}_{\tau\sigma}=\displaystyle\lim_{\Lambda \to 0}{E}_{\tau\sigma}^\Lambda$

\begin{figure}
\centering
\includegraphics[width=\textwidth]{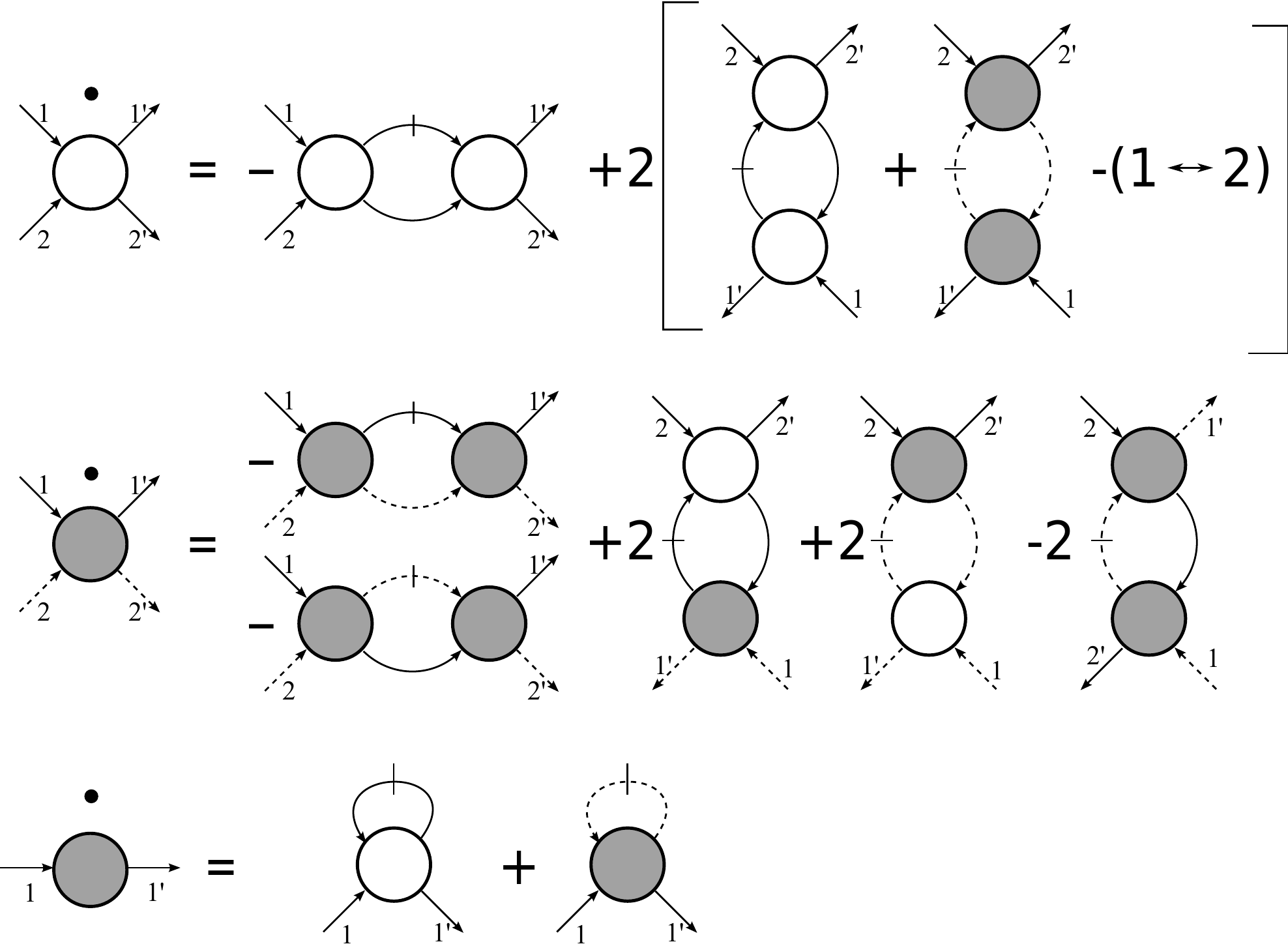}
\caption{Flow equation diagrams of the four-point vertices and the
self-energies. Dashed and full lines are the propagators
for the $\tau=A$ and $\tau=B$ orbitals, respectively.
White (filled) 4-armed circles are the intra-orbital (inter-orbital)
four-point vertices. Propagators with a vertical dash denote
single-scale propagators defined via Eq.~(\ref{eq:singlescale})
and the circular dot on the l.h.s. denotes differentiation with respect
to the cutoff $\Lambda$.}
\label{fig:diagrams}
\end{figure}
\subsection{Four-point vertices}

The frequency-independent
intra-orbital interaction is completely specified by the
interaction constants $U_A$ and $U_B$. Furthermore,
in absence of a magnetic field and of a direct hybridization of the
two orbitals, the $SU(2)$ spin symmetry allows
only two types of inter-orbital interaction: a charge interaction
which is specified by $U_{AB}$ and a Hund-type spin interaction
of magnitude $J$.
We can thus write the properly anti-symmetrized four-point vertex
as follows,
\begin{equation}
  \gamma_\Lambda^{(4)}
  =\gamma_\Lambda^{\rm inter}+\gamma_\Lambda^{\rm intra} \, ,
\end{equation}
with ($\delta_{\sigma_1\sigma_2}$ denote Kronecker deltas)
\numparts
  \begin{eqnarray}
    \gamma_\Lambda^{\rm inter} &=\sum_{\tau=A,B}
    U^\Lambda_\tau
    \delta_{\tau_1 \tau} \delta_{\tau_2 \tau}
    \delta_{\tau_1\tau_1'} \delta_{\tau_2\tau_2'}
     \nonumber
     \\ & \quad \times
    \big( \delta_{\sigma_1 \sigma_1'}
    \delta_{\sigma_2\sigma_2'} - \delta_{\sigma_1\sigma_2'}
    \delta_{\sigma_2\sigma_1'} \big)
    \label{eq:gaminter}
    \, , \\
    \gamma_\Lambda^{\rm intra}&=
    \big( \delta_{\tau_1 A}\delta_{\tau_2 B}+\delta_{\tau_1
      B}\delta_{\tau_2 A} \big)
    \Big\{ \frac{U^\Lambda_{AB}}{2}\big( \delta_{\sigma_1 \sigma_1'}
    \delta_{\sigma_2\sigma_2'}  \delta_{\tau_1 \tau_1'}
    \delta_{\tau_2\tau_2'}
    \nonumber \\
    & \quad
    - \delta_{\sigma_1\sigma_2'} \delta_{\sigma_2\sigma_1'}
    \delta_{\tau_1 \tau_2'} \delta_{\tau_2\tau_1'} \big)
    +\frac{J_\Lambda}{4}
          \big( {\bm \sigma}_{\sigma_1 \sigma_1'}
    \cdot {\bm\sigma}_{\sigma_2\sigma_2'}  \delta_{\tau_1 \tau_1'}
    \delta_{\tau_2\tau_2'}
     \nonumber   \\ & \quad
    - {\bm\sigma}_{\sigma_1\sigma_2'}\cdot
    {\bm \sigma}_{\sigma_2\sigma_1'}  \delta_{\tau_1 \tau_2'}
    \delta_{\tau_2\tau_1'} \big) \Big\} \, .
    \label{eq:gamintra}
      \end{eqnarray}
\endnumparts
where we dropped the arguments of the vertices, e.g. we implied
$\gamma_\Lambda^{\rm inter}=\gamma_\Lambda^{\rm inter}(1',2';2,1)$ etc.
These are all possible static two-particle correlations in presence of SU(2) symmetry
and for two orbitals, in absence of a single particle hybridization of the two dots.

\subsection{Flow equations}
Fig.~\ref{fig:diagrams} shows a diagrammatic representation of the flow
equations for the two-point and four-point vertices.
The flow equation of $E_{\tau\sigma}^\Lambda$ directly follows from
the flow equation of the self-energy Eq.~(\ref{eq:flowsigma}),
together with the form of the four-point vertex in Eqs.~(\ref{eq:gaminter})
and (\ref{eq:gamintra}).
We
keep here a $\sigma$-dependence of the self-energy which arises
in presence of a magnetic field.
We find
\begin{eqnarray}
  \partial_\Lambda E^\Lambda_{\tau\sigma} &=
  \frac{E^\Lambda_{\tau{\bar\sigma}} U^\Lambda_\tau/\pi}
  {(E^\Lambda_{\tau{\bar\sigma}})^2+(\Lambda+\Gamma_\tau)^2} \nonumber
  \\ & \quad +
  \frac{E^\Lambda_{\bar{\tau}\sigma} (U^\Lambda_{AB}+J_\Lambda/2)/(2\pi)}
  {(E^\Lambda_{\bar{\tau}\sigma})^2+(\Lambda+\Gamma_{\bar{\tau}})^2}
  \nonumber \\
  & \quad
  +
  \frac{E^\Lambda_{\bar{\tau} \bar\sigma} (U^\Lambda_{AB}-J_\Lambda/2)/(2\pi)}
  {(E^\Lambda_{\bar{\tau}\bar\sigma})^2+(\Lambda+\Gamma_{\bar{\tau}})^2} \, .
  \label{eq:flowV}
\end{eqnarray}
The flow equations of the coupling constants which parametrize the
four-point vertex follow from Eq.~(\ref{eq:flow4pnt}). For small
magnetic fields,
we neglect
the effect of the magnetic field on the four-point vertex, since
these yield corrections in the self-energy of only second
or higher order in the magnetic field.
We thus replace the $\sigma$-dependent energy levels
$E_{\tau\sigma}^\Lambda$ by the average value
$E_\tau^\Lambda=(E_{\tau\uparrow}^\Lambda+E_{\tau\downarrow}^\Lambda)/2$
in the flow
of the four-point vertex and arrive at
\numparts
  \begin{eqnarray}
     \partial_\Lambda U^\Lambda_\tau &=
     \frac{2\big(E^\Lambda_\tau U^\Lambda_\tau\big)^2
     }{\pi \big[(E^\Lambda_\tau)^2+
       (\Lambda+\Gamma_\tau)^2\big]^2}
     \nonumber \\
     & \quad
     + \frac{[(E^\Lambda_{\bar{\tau}})^2-(\Lambda+\Gamma_{\bar{\tau}})^2]
       [\frac{3}{4}{J_\Lambda}^2-(U^\Lambda_{AB})^2]}
     {2 \pi \big[(E^\Lambda_{\bar{\tau}} )^2
       +(\Lambda+\Gamma_{\bar{\tau}})^2\big]^2} \, ,
     \label{eq:flowUA}
     \\
     \partial_\Lambda U^\Lambda_{AB} &=
     -\sum_{\tau=A,B}\frac{\big[(E^\Lambda_\tau)^2
       -(\Lambda+\Gamma_\tau)^2\big]
       U^\Lambda_\tau U^\Lambda_{AB}}
     {\pi \big[(E^\Lambda_\tau)^2+(\Lambda+\Gamma_\tau)^2\big]^2}
     \nonumber \\
    &
    +\frac{[3E^\Lambda_A E^\Lambda_B-(\Lambda+\Gamma_A)(\Lambda+\Gamma_B)]
      [\frac{3}{4}{J_\Lambda}^2+(U^\Lambda_{AB})^2]}{2 \pi
      \big[(E^\Lambda_A)^2+(\Lambda+\Gamma_A)^2\big]
      \big[(E^\Lambda_B)^2+(\Lambda+\Gamma_B)^2\big]} \, ,
    \label{eq:flowUAB}
    \\
    \partial_\Lambda J_\Lambda  &=  \sum_{\tau=A,B}
    \frac{\big[(E_\tau^\Lambda)^2-(\Lambda+\Gamma_\tau)^2\big]
      U^\Lambda_\tau J_\Lambda}
     {\pi \big[(E^\Lambda_\tau)^2+(\Lambda+\Gamma_\tau)^2\big]^2}
    \nonumber \\
    &
    +\frac{2E^\Lambda_AE^\Lambda_B J_\Lambda U^\Lambda_{AB} -
      (\Lambda+\Gamma_A)(\Lambda+\Gamma_B){J_\Lambda}^2/2}
    {\pi \big[(E^\Lambda_A)^2+(\Lambda+\Gamma_A)^2\big]
      \big[(E^\Lambda_B)^2+(\Lambda+\Gamma_B)^2\big]} \, .
    \label{eq:flowJ}
  \end{eqnarray}
\endnumparts
 where $\bar{\tau}=B$ for $\tau=A$ and vice versa.

\section{Results}

We have solved the flow equations (\ref{eq:flowV}-\ref{eq:flowJ})
numerically, starting from a very large cutoff $\Lambda_0$ and
integrating the flow equations to $\Lambda=0$.
As discussed in Ref.~\cite{Meden06}, the appropriate
initial values at $\Lambda_0\to \infty$ of the flow parameters
are
\numparts
  \begin{eqnarray}
    E_{\tau\sigma}^{\Lambda_0}&= E_{\tau\sigma} \, , \\
    % \end{equation}
    % and $
    U_\tau^{\Lambda_0}&=U_\tau \, , \\
    U_{AB}^{\Lambda_0}&=U_{AB} \, , \, \\
    J_{\Lambda_0}&=J \, .
  \end{eqnarray}
\endnumparts
In particular, the bare one-particle contribution
arising from the $U_{AB}$ and $U_{A/B}$
terms in Eq.~(\ref{eq:Hd}) are exactly canceled when integrating the flow
equations from $\Lambda=\infty$ to a large but finite
$\Lambda_0$.\cite{Meden06}
We limit the discussion to cases with a ferromagnetic coupling $J<0$, for
which the singlet-triplet crossover can be observed.
While in most cases the flows are free of
divergences, for parameter sets corresponding
to strong triplet coupling, the interaction coupling constants diverge for $\Lambda\to 0$.
This divergence is probably an artefact of the approximation,
in which dynamic and three-particle
correlations are ignored. Similar divergences appear in some quantum dot
geometries even in absence of Hund's coupling terms.\cite{Weyrauch08}

\begin{figure}
\centering
\includegraphics[width=\linewidth]{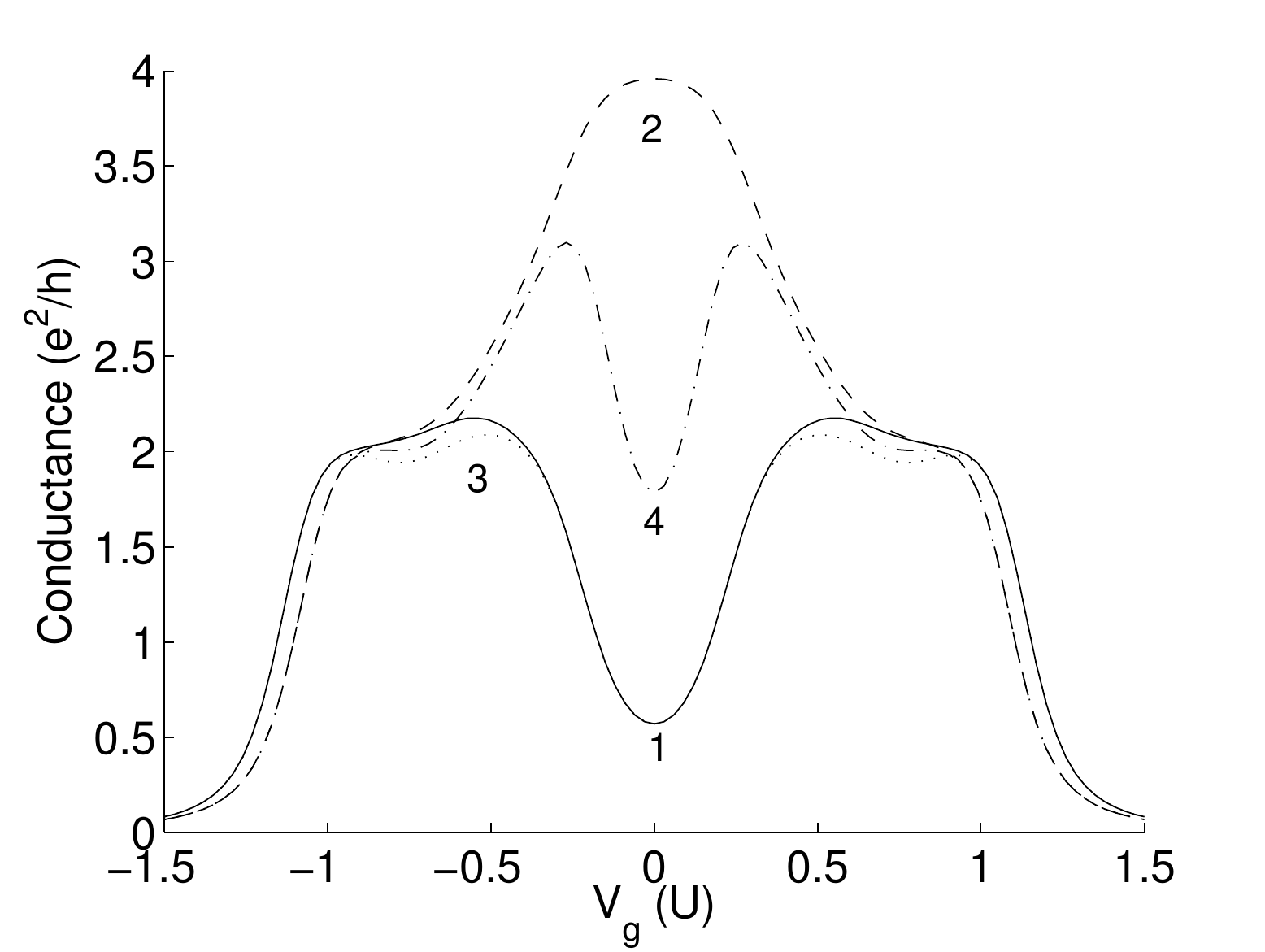}
\caption{Conductance as a function of the gate voltage $V_g$ for
$U_A=U_B=U_{AB}=U$, $J=-0.3U$, $\Gamma_{A}=\Gamma_B=0.02\pi U$.
A small change in level separation pushes the system into the triplet configuration. (Curves 1 and 2, $\delta=0.5U,0.4U$, respectively.) Adding a small magnetic field, one sees that the triplet conduction peak is much more fragile than the regular Kondo $S=1/2$ plateaus. ($1\rightarrow 3$, $2\rightarrow 4$.)}
\label{fig:singlettriplet}
\end{figure}

Fig.~\ref{fig:singlettriplet} shows the dependence of the conductance
for two different values of the level splitting, $\delta=0.5U$ (curve 1)
and $\delta=0.4U$ (curve 2),
as a function of the gate voltage $V_g$ which shifts the energy
levels according to Eqs.~(\ref{eq:levpos}). For large $\delta$ one observes
two well separated conductance plateaus which each arise from the usual
spin-$1/2$ Kondo effect. In the first plateau region the occupation of the
lowest
orbital is one while the upper orbital is empty. In the second plateau region
the lowest orbital is fully occupied whereas the upper orbital is singly occupied. Around $V_g=0$ the occupancy of the lower orbital is close to two
(singlet)
whereas the upper orbital is almost empty. This leads to a strongly
suppressed conductance at and around $V_g=0$.
A slightly smaller value of the level splitting $\delta$ induces
a singlet-triplet transition which
leads to the
occurrence of a conductance plateau with $G\approx 4 e^2 /h $ around $V_g=0$.

While our approach does not allow to calculate the temperature dependence of the
conductance, we can investigate its field dependence (at zero temperature) to study the stability
of the observed conductance plateaus since
a small magnetic field suppresses the Kondo effect in a smilar manner as a
finite temperature.
The peak of the conductance at $V_g=0$ is clearly rather fragile,
a small magnetic field cuts the peak in half (curve 4) while the regular $S=1/2$ Kondo plateau hardly changes for comparable values of the magnetic field (curve 3). We can in fact define a Kondo scale as e.g. the value of Zeeman splitting $h$ produced by a magnetic
field necessary to suppress conductance by a factor of 1/2.

If the level
separation is made very small and close to $V_g \approx 0$,
eventually all coupling
constants diverge when $\Lambda$ approaches zero. The divergence
of the parameters,
which would presumably disappear in a more refined approximation,\cite{Weyrauch08}
is however such that they define a triplet state.

\begin{figure}
\centering
\includegraphics[width=\linewidth]{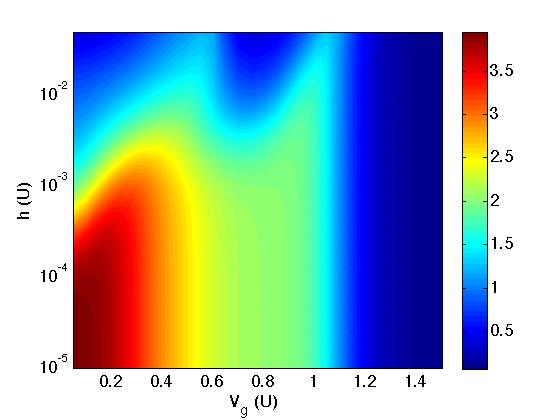}
\caption{Conductance as a function of gate voltage $V_g$ and magnetic field splitting $h$, $\delta=0.4U$. Here it is very clear how the singlet-triplet plateau is suppressed at smaller $h$ than the $S=1/2$ Kondo plateau.
}
\label{fig:singlettriplet2}
\end{figure}

As we already mentioned, one of the main advantages of FRG over NRG is the ability to reliably calculate zero-frequency properties very quickly. One can thus scan large parts of the parameter space to produce 2D maps such as the one in Fig.~\ref{fig:singlettriplet2}. It shows the conductance as a function of $V_g$ and Zeeman splitting $h$. Here, one sees again how the triplet plateau is suppressed at a lower value of $h$ than the $S=1/2$ plateau. Since $U_A=U_B$, the conductance is symmetric around $V_g=0$ and it is thus sufficient to plot for $V_g>0$.

\begin{figure}
\centering
\includegraphics[width=\linewidth]{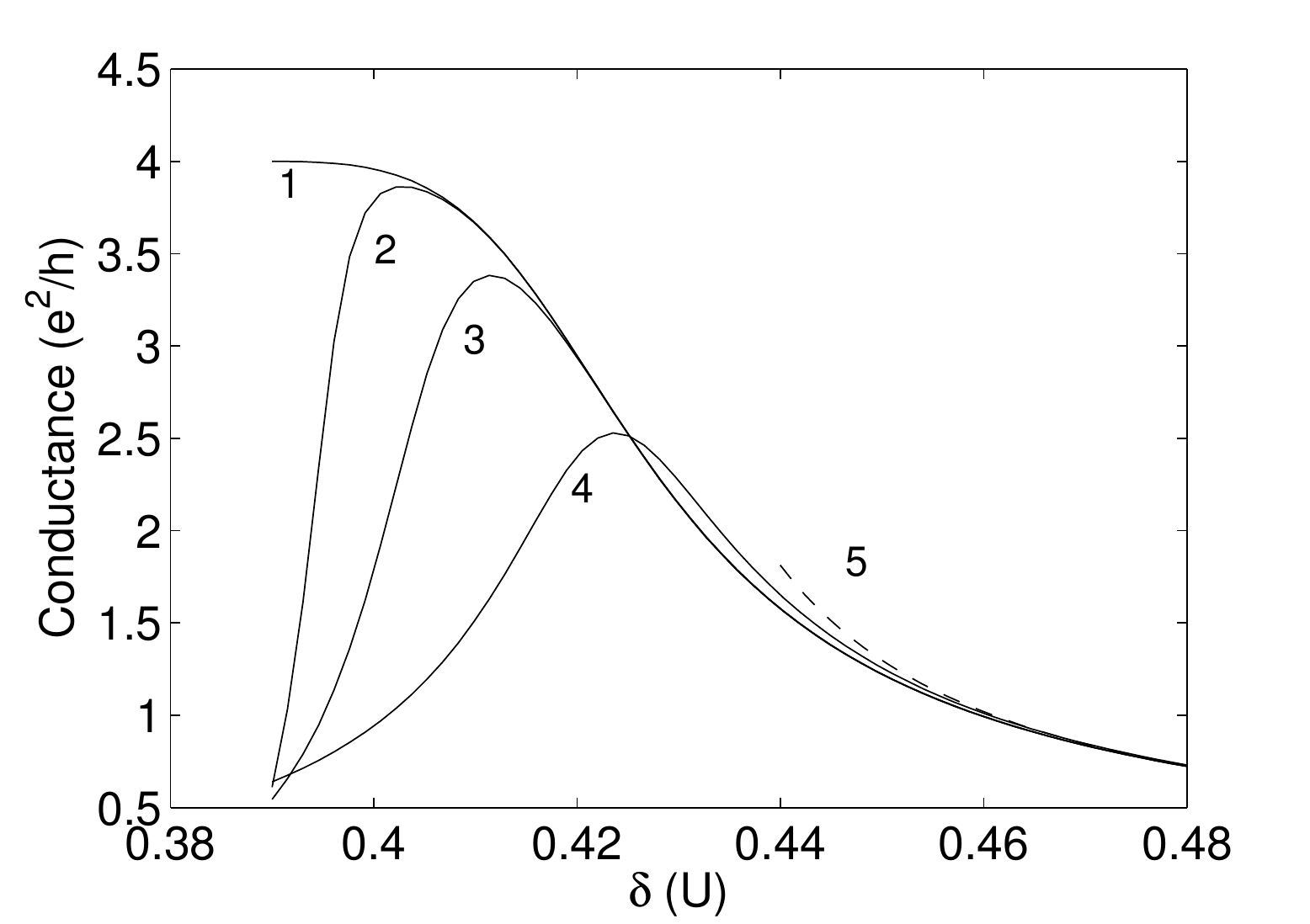}
\caption{Singlet-triplet transition for $h/U=0, 10^{-4}, 10^{-3}, 8\times10^{-3}$ (curves 1-4). Dashed curve (5) is a fit according to Eq.~(\ref{eq:asymptote}).
}
\label{fig:transition}
\end{figure}

\begin{figure}
\centering
\includegraphics[width=\linewidth]{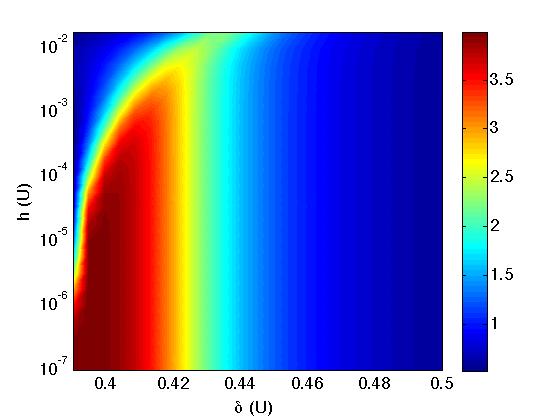}
\caption{Singlet-triplet transition as a function of $h$. One can see how the conduction on the triplet side gets suppressed and a peak at the singlet-triplet transition remains. The swerve of the peak to higher $\delta$ at the top of the plot is because the magnetic splitting $h$ is becoming comparable to the $x$-axis unit.
}
\label{fig:transition2}
\end{figure}

Figure \ref{fig:transition} shows the conduction at the symmetry point
$V_g=0$ as a function of the level separation $\delta$, for different
values of magnetic field $H$. As seen by the curve corresponding to $H=0$ (curve 1),
this regime of $\delta$ covers the singlet-triplet transition.
Even a very small magnetic field suppresses the conduction on the triplet
side strongly, but at the transition, a conduction peak remains.
The dashed line is an asymptote taken from Eq.~(\ref{eq:asymptote}),
valid for large $\delta$. The derivation of the asymptotic result
Eq.~(\ref{eq:asymptote}) does however only take into account
the triplet and local singlet configuration of the quantum dot
and becomes inaccurate for too large $\delta$. As before, we also plot the results on a 2D map, shown in Fig.~\ref{fig:transition2}.

\begin{figure}
\centering
\includegraphics[width=\linewidth]{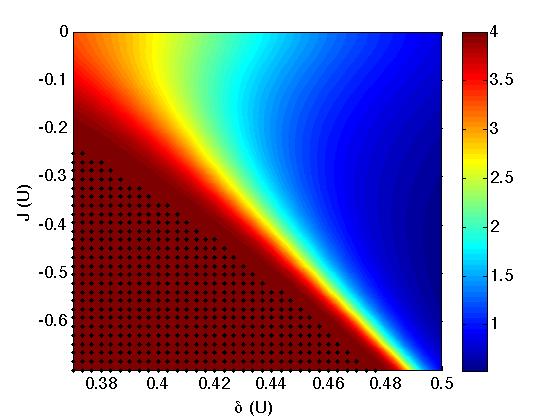}
\caption{Conductance as a function of $\delta$ and $J$. Direct Coulomb interactions are the standard ones, and we are at the symmetry point $V_g=0$. As expected from Eq.~\eref{eq:STcondition}, there is a linear relation between $\delta$ and $J$ along the singlet-triplet transition curve. However, renormalization effects shift the line such that the slope is not $1/4$ and it does not go through $(\delta,J)=(0,0)$. The slope is about $0.2$ and an extrapolation shows that the curve passes through $J=0$ at $\delta\simeq 0.37U$. The black dots correspond to parameters where the flow diverges.}
\label{fig:scan_d_J}
\end{figure}

\begin{figure}
\centering
\includegraphics[width=\linewidth]{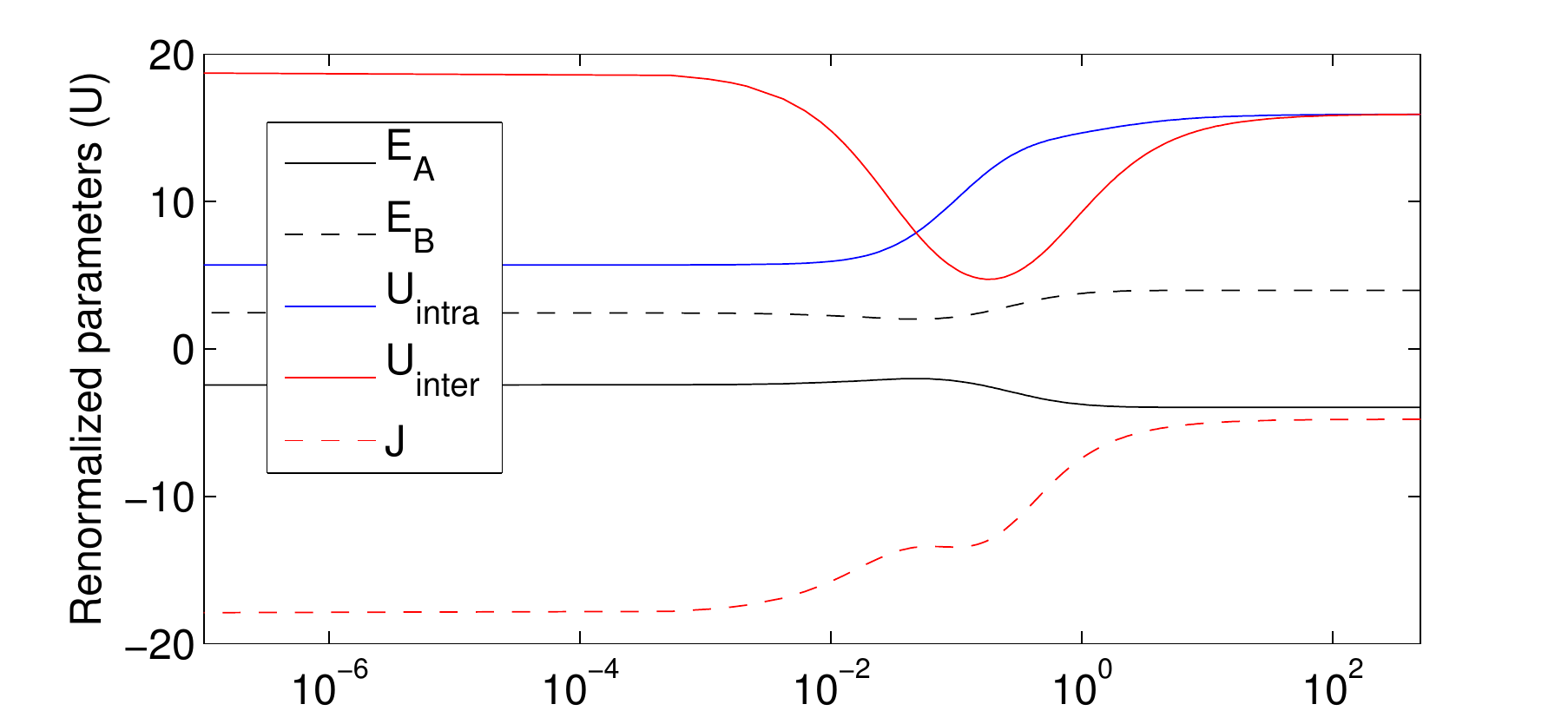} \\
\includegraphics[width=\linewidth]{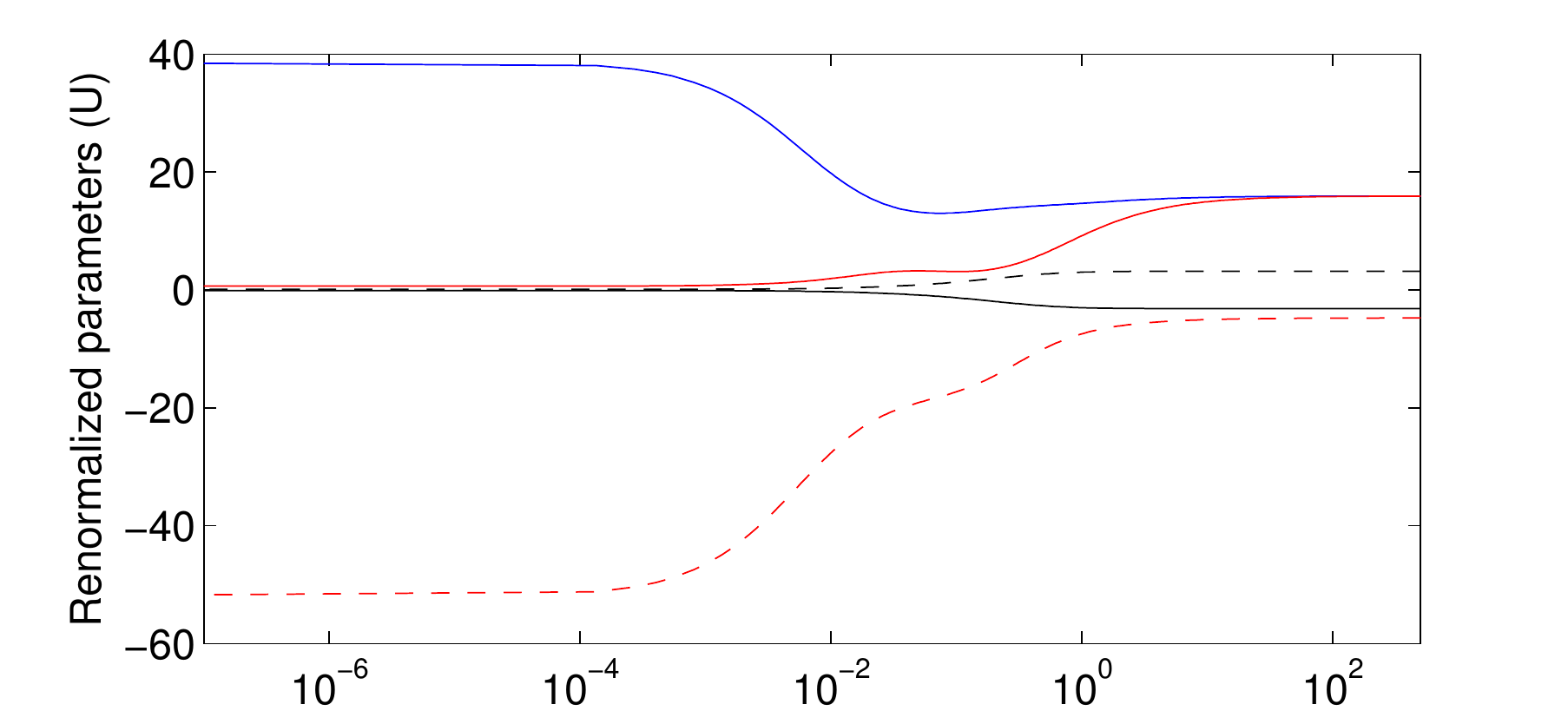} \\
\includegraphics[width=\linewidth]{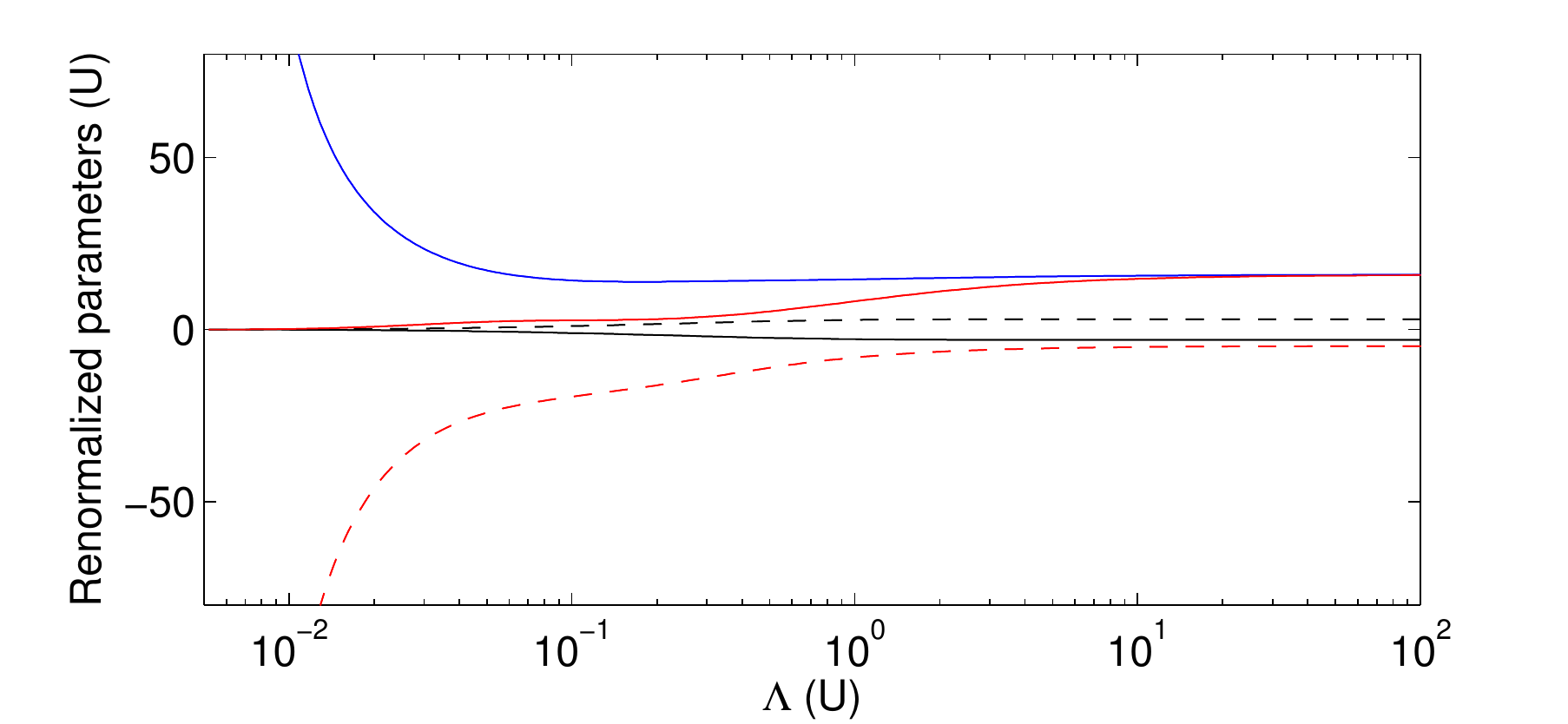}
\caption{Flow of parameters for $U_A=U_B=U_{AB}=U$, $J=-0.3U$, $\Gamma_{A}=\Gamma_B=0.02\pi U$, $V_g=0$. From top to bottom, $\delta=0.5U, 0.4U, 0.38U$. Top: final value of the inter-orbital Coulomb repulsion is much larger than the intra-orbital one, so it the electrons both stay on the bottom level in the singlet configuration. Middle: The inter-orbital Coulomb repulsion vanishes and the Hund's exchange grows very large, so the triplet configuration becomes favorable. Bottom: at very small level separation, the flow diverges at some value of $\Lambda$, but at least one can see from the divergence that the system is in the triplet state.
}
\label{fig:flow}
\end{figure}

To further illustrate the aforementioned triplet divergence of parameters, we plot the conductance for $V_g=0$ as a function of $\delta$ and $J$ in Fig.~\ref{fig:scan_d_J}. The black dots mark spots where the flow diverged, clearly showing that this occurs when the system is strongly defined to be in the triplet state, with small $\delta$ and large $|J|$. The flow of the parameters is visualized in Fig.~\ref{fig:flow} for three sets of initial parameters. The top plot shows the flow on the singlet side of the transition, the middle one on the triplet side but still close to the transition, and the bottom one far into the triplet side, where the divergence occurs. But, as stated before, the parameters diverge in a controlled way: $|J|$ and $U_A,U_B$ become very large ($J$ stays negative) while $U_AB$ goes to zero, so it is clearly preferable to have the two electrons in the triplet state.

\section{Summary and conclusions}
In summary, we have extended the static FRG approach presented in
Ref.~\cite{Meden06} for multi-orbital quantum dots
to include also Hund's coupling and studied the triplet-singlet
transition in vertically coupled dots, at $T=0$. Despite its simplicity,
the here presented approach reproduces qualitatively the low-temperature
results from NRG approaches.\cite{Izumida01} The dependence of
the conductance on the gate voltage follows the predictions
of the perturbative RG analysis \cite{Glazman88} near the triplet-singlet transition.

Higher order static correlations would probably give rise to
small quantitative improvements, but a discussion of the full
temperature and magnetic field dependence of the conductance would
require an extension to a dynamic approximation, which, at the moment,
are only available for the weak to moderate coupling regime.\cite{Karrasch08,Bartosch09,Isidori10}

 N.H. thanks Aldo Isodori and Johannes Bauer for discussions. We acknowledge support from the FP7 IRSES "SPINMET" grant. E.~B.~M.~acknowledges support from Rannis, the Icelandic Centre for Research. NH further acknowledges support from
the DFG-CNPq project 444BRA-113/57/0-1 and the DFG research
group FOR 723.


\begin{thebibliography}{99}

\bibitem{Glazman88} L.~I.~Glazman and M.~E.~Raikh,
  JETP Lett. {\bf 47}, 452 (1988); T.~K.~Ng and P.~A.~Lee,
  Phys.~Rev.~Lett. {\bf 61}, 1768 (1988).
\bibitem{Meir02} Y.~Meir, K.~Hirose and N.S.~Wingreen,  Phys.~Rev.~Lett. {\bf 89}, 196802 (2002).
\bibitem{Sasaki01}S. Sasaki, S. De Franceschi, J. M. Elzerman,
W. G. van der Wiel, M. Eto, S. Tarucha, and L. P. Kouwenhoven
Nature {\bf 405}, 764 (2000).
\bibitem{Kouwenhoven01} L.~P.~Kouwenhoven, D.~G.~Austing,
and S.~Tarucha, Rep-~Prog.~Phys. {\bf 64}, 701 (2001).
\bibitem{Florens11} S.~Florens et al., J.~Phys.: Condens.~Matter {\bf 23}, 
243202 (2011)
 \bibitem{Bulla08} R.~Bulla, T.~A.~Costi, and T.~Pruschke,
  Rev.~Mod.~Phys. {\bf 80}, 395 (2008).
 \bibitem{Anders08} F.~Anders et al., Phys.~Rev.~Lett.~{\bf 100}, 086809 (2008)
\bibitem{Izumida01} W.~Izumida, O.~Sakai, and S.~Tarucha,
  Phys.~Rev.~Lett. {\bf 87}, 216803 (2001);
  W.~Izumida and O.~Sakai, J.~Phys.~Soc.~Jpn.
  {\bf 74}, 103 (2005).

\bibitem{Hofstetter02} W.~Hofstetter and H.~Schoeller,
  Phys.~Rev.~Lett. {\bf 88}, 016803 (2001).
\bibitem{Pustilnik03} M.~Pustilnik, L.~I.~Glazman, and W.~Hofstetter,
Phys.~Rev.~B {\bf 68}, R161303 (2003).
\bibitem{Hofstetter04} W.~Hofstetter and G.~Zarand, Phys.~Rev.~B
  {\bf 69}, 235301 (2004).
\bibitem{Karrasch08} C.~Karrasch, R.~Hedden, R.~Peters, T.~Pruschke,
K.~Sch\"onhammer, and V. Meden, J. Phys.: Condens. Matter
{\bf 20}, 345205 (2008).
\bibitem{Bartosch09}
L. Bartosch, H. Freire, J.~J.~R.~Cardenas, and P.~Kopietz,
J.~Phys.~Condens.~Matter {\bf 21}, 305602 (2009).
\bibitem{Isidori10} A.~Isidori, D.~Roosen, L.~Bartosch, W. Hofstetter, and
P.~Kopietz,  Phys.~Rev.~B {\bf 81}, 235120 (2010).
\bibitem{Meden06} C.~Karrasch, T.~Enss and V.~Meden, Phys.~Rev.~B
  {\bf 73}, 235337  (2006).
\bibitem{Horvath10} B.~Horv\'ath, B.~Lazarovitz, and G.~Zar\'and,
Phys. Rev. B 84, 205117 (2011).

\bibitem{Hermann} H. Freire and E. Corr\^ea, J. Low. Temp. Phys. 166, 192 (2012).

\bibitem{Pustilnik01} M.~Pustilnik and L.I.~Glazman,
  Phys.~Rev.~Lett. {\bf 85}, 2993 (2000);
  Phys.~Rev.~B 64, 045328 (2001).
\bibitem{Meir92}
 Y.~Meir and N.~S.~Wingreen, Phys.~Rev.~Lett.~{\bf 68}, 2512 (1992).
\bibitem{Wetterich93} C. Wetterich, Phys. Lett. B {\bf 301}, 90 (1993).
\bibitem{Morris94} T. R. Morris, Int. J. Mod. Phys. A {\bf 9}, 2411 (1994).
\bibitem{Metzner11} W.~Metzner, M.~Salmhofer, C.~Honerkamp, V.~Meden, and
  K.~Sch\"onhammer, Rev. Mod. Phys. {\bf 84}, 299 (2012); P.~Kopietz, L.~Bartosch, and F.~Sch\"utz, ``Introduction
to the Functional Renormalization Group'', Springer (2010).
\bibitem{Weyrauch08} M.~Weyrauch and D.~Sibold , {Phys. Rev. B} {\bf 77}, 125309 (2008).

\end{thebibliography}
\end{document}